\documentclass[12pt]{article}

\usepackage{dcolumn}
\usepackage{bm}
\usepackage[latin1]{inputenc}
\usepackage[spanish,english]{babel}
\usepackage{amsfonts}
\usepackage{amssymb}
\usepackage{graphicx}
 \usepackage{setspace}
 \usepackage{verbatim} 

\usepackage{amsmath}
\numberwithin{equation}{section}

\usepackage{color}
\usepackage{soul} 
 
\usepackage{slashed} 
 
 \usepackage{ulem}

\usepackage[margin=3cm]{geometry}

\newcommand{\be}{\begin{equation}}
\newcommand{\ee}{\end{equation}}
\newcommand{\bea}{\begin{eqnarray}}
\newcommand{\eea}{\end{eqnarray}}

\begin{document}

\begin{titlepage}

\begin{center}
{\large \bf 
Black holes, conformal symmetry,  and fundamental fields\\
} 
\end{center}

\begin{center}

Jos\'e Navarro-Salas
\\

{  \footnotesize \noindent {\it Departamento de F\'isica Te\'orica and IFIC, Universidad de Valencia-CSIC,  E-46100, Burjassot (Valencia) Spain. \\ Email: jnavarro@ific.uv.es
} }



\end{center}


\begin{abstract}

 Cosmic censorship protects the outside world from black hole singularities and paves the way for assigning entropy to gravity at the event horizons. We point out a tension between cosmic censorship and the quantum backreacted geometry of  Schwarzschild black holes, induced by vacuum polarization and driven by the conformal anomaly. A similar tension appears for the Weyl curvature hypothesis at the Big Bang singularity. We argue that the  requirement of  exact conformal symmetry resolves both conflicts and has major implications for constraining the set of 
fundamental constituents of the Standard Model.

\end{abstract}

\vspace{3cm}

\end{titlepage}

\section{Introduction}
\setcounter{equation}{0}
General relativity predicts two different types of spacetime singularities: inside black holes  and at the beginning of time \cite{Penrose65, HP70}. 
Penrose's cosmic censorship \cite{Penrose69, Penrose73} shields 
the outside world from black hole singularities, which have a divergent Weyl tensor  hidden from outside observers by event horizons. This prevents something worse (naked singularities) from happening, 
whereas 
the  Weyl curvature hypothesis (WCH) \cite{Penrose79}, in the form given in \cite{Tod03}, protects the smoothness of the Big Bang singularity, which is assumed to be  conformally regular.   
These issues have been traditionally dissociated in  discussions of the specific and detailed structure of the Standard Model of particle physics \cite{Schwartz}.

The aim of this work is to relate the cosmic  censorship/WCH to fundamental quantized fields.
The link between both areas  is based  on local conformal symmetry, defined by the invariance under Weyl transformations of the metric tensor
\be \label{1} g_{\mu\nu}(x) \to \Omega^2(x) g_{\mu\nu}(x) \ . \ee 
In a first approximation, ignoring interactions and masses, the Standard Model is a conformally invariant theory. 
 We will try to consistently combine concepts related to general relativity, in particular the emergence of spacetime singularities, with the properties of the field content of the Standard Model. 

 On a primary level, we can assume the 
 semiclassical Einstein equations 
\be \label{Esemi}R_{\mu\nu} - \frac{1}{2} R g_{\mu\nu} = 8\pi G \langle T_{\mu\nu} \rangle \ , \ee
and the quantum properties of the (renormalized) stress-energy tensor $\langle T_{\mu\nu} \rangle$ of matter fields. We will assume a leading order approximation, so the classical stress-energy tensor of the  fundamental fields is traceless, as a consequence of  conformal invariance. 
The first relevant novelty in solving (\ref{Esemi}) is that the trace of $\langle T_{\mu\nu} \rangle$ is not zero. This is due to the breaking of local conformal symmetry when quantum free  fields couple to gravity \cite{CD, DDI}, as described in the textbooks \cite{birrell-davies, parker-toms, hu-verdaguer}. The quantum trace is given by a combination of geometric tensors \cite{Duff}
\be \label{4danomaly0} \langle  T^{\mu}_\mu  \rangle = \hbar (c\ C^2 -a \ E + d \ \Box R )\ , \ee
 where $C^2$ is the square of the Weyl curvature, and $E$ is the Euler density. The constant coefficients $c, a$ are univocally  fixed by renormalization and depend  on the spin of the field \footnote {The coefficient $d$ is generally ambiguous. In addition to the spin dependence, it also depends on the renormalization scheme.}.

To display our arguments we will first consider the structure of the semiclassical Schwarzschild geometry. 
In Section \ref{bh} we will describe the results in the literature for the semiclassical static corrections. Despite the assumption of staticity, this is still a very challenging problem. However, the conformal behavior of the semiclassical Schwarzschild spacetime can be captured with some approximation techniques. The main feature is that the semiclassical geometry  is horizonless. It has a curvature singularity without being surrounded by a horizon. We will argue that the driving force leading to this geometry is the existence of a non-vanishing trace anomaly.  A   useful 
interpretation of this result is that the gravitational vacuum polarization  may be strong enough to allow for ultracompact stellar configurations whose external geometry is described by a (horizonless) portion of the  semiclassical Schwarzschild geometry.  This is indeed a very interesting area of research (see, for instance, the review \cite{Cardoso-Pani}).

 In this work  we want to point out  an alternative  possibility. 
Our reasoning is described in Section \ref{Tcancellation}.  The alternative and more risky option is to demand that the fundamental constituents of the Standard Model, or some appropriate extension of it, conspire to cancel out the conformal anomaly. 
This condition is not trivial, since the numerical coefficients in (\ref{4danomaly0}) involve fractions of integers. The vanishing of the total trace anomaly involves a very minimal extension of the Standard Model. In addition to the known particles, it requires  of three generations of  right-handed  neutrinos. Fortunately, the numerology for the cancellation of the conformal anomaly has recently been
reported in \cite{BT21, MVZ}, and it requires  conformally invariant fields of zero dimension to fix the vacuum. No particle excitations are allowed for these unconventional scalar fields. 

In Section 4 we consider the implications of exact conformal invariance for the Big Bang singularity. We argue that it is fully consistent with the WCH, thus strengthening  the main message of this work: exact conformal invariance paves the way for protecting physics from  spacetime singularities.  As a bonus, we obtain a highly non-trivial constraint on the field content of the Standard Model. 

The right-handed neutrinos play a crucial role for the simplest explanation of the observed oscillations of the left-handed neutrinos, via the seesaw mechanism \cite{Minkowski, SM}. Here the sterile neutrinos are also required for exact conformal symmetry. Furthermore,  the leading correction in the massive sector, according to the seesaw mechanism, comes from the heavy right-handed neutrinos. The addition of a (Majorana) mass term to a fermionic field has a major implication. The expansion of the universe creates particles out of the vacuum \cite{parkerthesis, Parker68, Parker69, Parker71}. For spin $1/2$ fields with zero mass, the expansion does not produce particles due to the underlying conformal invariance of the field equations. The expected heavy mass of the sterile fermion, which only interacts with gravity, is therefore the source of its own creation from the vacuum. A robust calculation of the particle production rate at late cosmic times requires a characterization of the initial vacuum at the Big Bang event. Going back to the Big Bang requires a quantum theory of gravity. This is a very hard problem. However, from a phenomenological point of view, the approach to the Big Bang can be parameterized by a temporal window function. It is a mathematical tool of quantum field theory in curved spacetime to construct a low-energy vacuum state that satisfies the Hadamard condition
 \cite{Olbermann, BN, NNP23}. 
This point is also  discussed in Section 4. We also include an estimate of the mass of the heavy right-handed neutrino that is consistent with the observed abundance of dark matter.

Finally, in Section 5, we summarize our conclusions. We would like to note that this work should be seen as an attempt to harmonize ideas from different camps.  It has been inspired by recent work that re-emphasizes  the expected fundamental role of local conformal invariance \cite{Hooft} in understanding fundamental aspects of the microuniverse 
as well as its large-scale structure.

\section{Black holes, vacuum polarization, and cosmic censhorship} \label{bh}

Let us first recall the well-known  description of the Schwarzschild black hole, which is considered to be the final (stationary) state of a  spinless gravitational collapse.
A generic 
spherically symmetric metric can be expressed as 
\be  ds^2 =  g_{ab}dx^a dx^b + r^2 d\Omega \ . \ee
When  the classical vacuum Einstein field equations are imposed, one obtains the  solution
\be g_{ab}dx^a dx^b = -(1-2M/r)dt^2 + (1-2M/r)^{-1}dr^2 \ . \ee 
The curvature singularity at $r=0$ is hidden by the event horizon at $r=2M$.
Things get much more complicated when a quantum field  is added to the problem. 
The vacuum expectation values of the stress-energy tensor of a set of conformal  fields in the static
vacuum state   behave \cite{Candelas80} (we use geometrized units with $G=1=c$)
  
  \begin{equation} \label{Tcandelas}
\langle T^{\mu}_{ \ \nu} \rangle \sim _{r \to 2M} - \frac{\left(2 N_1+\frac{7}{2} N_{1 / 2}+N_0\right)}{30 \ 2^{12} \pi^2 M^4} \frac{\hbar }{ (1-2M/r)^2}\left(\begin{array}{cccc}
-1 & 0 & 0 & 0 \\
0 & \frac{1}{3} & 0 & 0 \\
0 & 0 & \frac{1}{3} & 0 \\
0 & 0 & 0 & \frac{1}{3}
\end{array}\right)
 \ .  \end{equation}
  The vacuum   has been defined by requiring the
normal modes to be positive frequency
with respect to the Killing vector $\partial_t$ with respect
to which the Schwarzschild exterior region is static \cite{Boulware}.
  $N_0$, $N_{1}$, and $N_{1/2}$ are integer numbers that  refers to the number of spin-$0$, spin-$1$ and Weyl  spinor fields, respectively. $M$ is the mass of the black hole.

We observe that the static state leads to  a  divergence of the stress-energy tensor at the horizon of the classical black hole background. 
 This is the main argument often used  to exclude the horizon as a physical portion of the Boulware state.
However, the above claim is circular in the sense that it assumes the existence
of a horizon that persists in the backreacted geometry obtained by solving the semiclassical Einstein equations. 
Therefore, in order to make progress, we must  study the backreaction problem. 
In this state, quantum
fields will generally exhibit  vacuum polarization, i.e., the expectation value of the stress-energy
tensor $\langle T_{\mu\nu} \rangle$ will be non-vanishing, as can be seen from the result (\ref{Tcandelas}).
However, solving the semiclassical Einstein equations  is a daunting task because we do not have sufficient analytical control over the general vacuum expectation values
$$ \langle T_{\mu\nu} (g_{\rho\sigma})\rangle \ . $$
This has been a topic of technical research in the literature for decades, with increasing efficiency \cite{Howard-Candelas,
Anderson-Hiscock-Samuel, Levi-Ori, TBO}.

 \subsection{Two-dimensional trace anomaly and the semiclassical \\ Schwarzschild geometry}

Due to the technical difficulty of the problem, one can expect to gain insight into the global structure of the backreaction geometry by making useful approximations from the outset. 
Consider for example a massless scalar field $\varphi$ and, for simplicity, restrict its expansion in spherical harmonics to the $s$-wave (spherical symmetry). The truncated theory can be described as effectively living in the $(t-r)$ sector, with $ \varphi = \varphi (x^a) $, and with action proportional to

\be \int d^2x \sqrt{-g^{(2)}} r^2 \ g^{ab} \partial_a \varphi \partial_b \varphi \nonumber 
\ . \ee
This action is invariant under conformal transformations of the  two-dimensional metric $g_{ab}$. The non-angular components of the stress-energy tensor are given by

\be  T_{ab} = \frac{T_{ab}^{(2)}}{4\pi r^2} \ , \ee
where 
\be T_{ab}^{(2)}= (\nabla_a \varphi \nabla_b \varphi-\frac{1}{2} g_{a b}(\nabla \varphi)^2) \  \ee
is the effective 2d stress-energy tensor. 

The backreacted geometry can be obtained, in a first approximation very popular in the 90's \cite{Harvey, Strominger}, by solving  the semiclassical Einstein equations  sourced  by the effective 2d stress-energy tensor $\langle T_{ab}^{(2)}\rangle$. The quantized theory does not respect the classical conformal symmetry, giving rise to a quantum anomaly. The most relevant part of  $\langle T_{ab}^{(2)}\rangle$ is determined by the corresponding conformal/trace anomaly, which is given by \cite{Mukhanov94, FN05}

\be \label{2danomaly}\langle T^{(2)a}_{\ \ a} \rangle = \frac{C \hbar }{24\pi} [R^{(2)} -6(\nabla \phi)^2 + 6\Box \phi]  \ , \ee
where $r^2 \equiv l^2 e^{-2\phi}$.  We note that $\phi$, usually referred in the two-dimensional literature as the dilaton field \cite{Harvey, Strominger, FN05}, is closely related to the usual radial function. The  radial coordinate is now considered a field $r=r(x^a)$. [$l$ is an arbitrary constant that has no role in the rest of the discussion and will be neglected. ]  The constant $C$ takes the unit value of $C=1$ for a single scalar field.

\begin{figure} \begin{center}\includegraphics[angle=0, width=60mm]{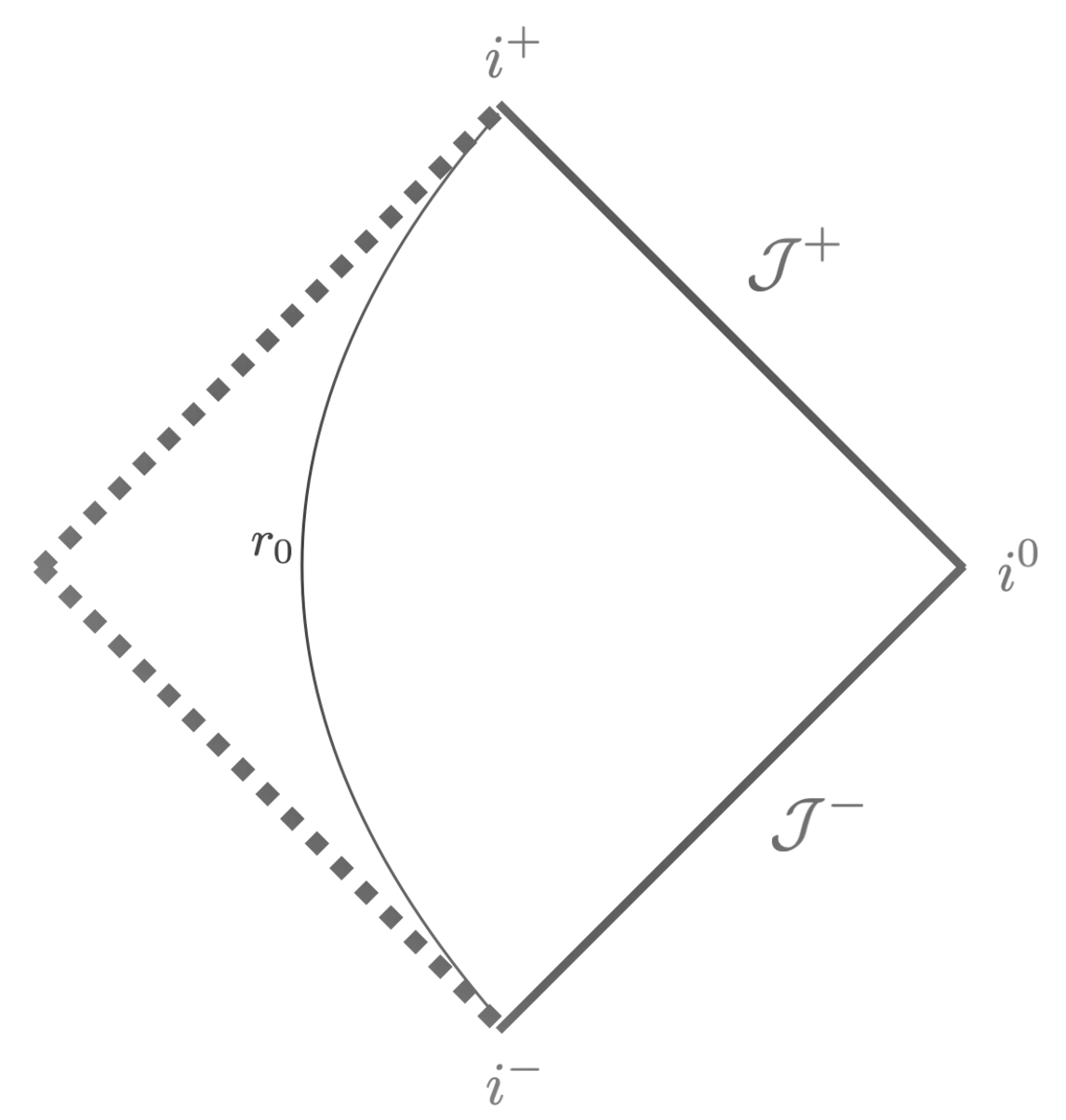}\end{center}
\caption{Penrose diagram showing the conformal structure of the semiclassical Schwarzschild geometry. The bouncing surface for the radial function is located at $r=r_0$.  
It is possible for an observer to approach the singularity and escape far away.}\label{censorship}\end{figure}

 In the near-horizon region $r \approx 2M$ we have $\nabla \phi \approx 0$, and one easily recovers the traditional trace anomaly of two-dimensional conformal field theory $\langle T^{(2)a}_{\ \ a} \rangle = \frac{C \hbar }{24\pi} R^{(2)}$, where $C$ is called the central charge \cite{CFT}. 
In this approximation (usually refereed to as the Polyakov approximation \cite{FN05}) the trace anomaly is sufficient to fix $\langle T_{ab}^{(2)} \rangle$. Introducing null coordinates and writing 
\be g_{ab}dx^adx^b =  -e^{2\rho} dx^+dx^- \ , \ee
the trace anomaly  together with the conservation equation $\nabla^a \langle T_{ab}^{(2)} \rangle=0$  determine 
$\langle T_{ab}^{(2)} \rangle$, up to two chiral functions $t_{\pm}(x^{\pm})$ \cite{Harvey,FN05}

\bea
\langle T_{ \pm \pm}\rangle=-\frac{C \hbar }{12 \pi}\left(\partial_{ \pm} \rho \partial_{ \pm} \rho-\partial_{ \pm}^2 \rho + t _{\pm}(x^{\pm}) \right) \ , \\ 
\langle T_{+-} \rangle = -\frac{C \hbar }{12 \pi}\partial_+ \partial_-\rho \ . 
 \eea
The functions $t_{\pm}(x^{\pm})$ parametrize the choice of the vacuum state. The static vacuum is characterized by the simplest choice  $t_{\pm}=0$. 
We have all the ingredients to obtain the backreacted Schwarzschild geometry (in the s-wave approximation) in the static coordinates $x^a=(t, x)$. The semiclassical Einstein equations reduce to a nonlinear ordinary differential equation. The technical details are not relevant and  can be found in \cite{FN06}.
The semiclassical solution is similar to the classical Schwarzschild solution until very close to the would be event horizon, which is now replaced by a  bouncing surface for the radial coordinate at $r=r_0$ ($\frac{dr}{dx}(r=r_0)=0$), mimicking the  throat of a  wormhole.
The backreacted metric has a  curvature singularity  that is not enclosed by a horizon 
\cite{FN06, Julio20, Ho}. 
The  resulting geometry  has the properties of a non-symmetric wormhole. An asymptotically flat branch connects the throat, and a null singularity develops beyond it. The classical horizon has been removed by quantum effects and a naked singularity emerges. Note that it is possible for an observer to get as close to the singularity as he/she wants and still send messages out to infinity. 
 The conformal structure of the semiclassical static solution is shown in  Figure \ref{censorship}.

This picture, based on  the effective two-dimensional conformal anomaly, can be interpreted as 
a  breakdown of cosmic censorship due to purely quantum effects.  
It also  anticipates the  result derived  from an intrinsic four-dimensional approach. At this point, it is interesting to pause for a moment and  try to implement the strategy mentioned in the introduction.
A natural way out of this tension (i.e., the breakdown of  the cosmic censorship induced by the two-dimensional trace anomaly)
is to realize that the physical effects of vacuum polarization are not  limited  to a single type of matter field.  We should sum up the contributions to the trace anomaly of all physically sensible (conformal) matter fields. However,  
unitarity imposes severe restrictions \cite{FQS84}, in particular $C \geq 0$ (for example, $C=1/2$ for fermions). This condition implies that
there is no consistent solution to the requirement that the sum of the central charges of the different conformal matter fields be zero. An overall cancellation is not possible, since there are no conformal matter fields with negative values for $C$ in (\ref{2danomaly}) \footnote{A negative value for $C$ arises from the unphysical diffeomorphism ghosts of the worldsheet in string theory. The cancellation of the trace anomaly determines the critical dimension \cite{P}.}. As we will  see, the problem changes drastically in a purely four-dimensional approach, where the degrees of freedom of the matter fields are not truncated to the $s$-wave modes.  

Before doing so, it is  worth briefly describing the situation in $1+2$  dimensions.
Quantum effects only lead  to a growth of the classical event horizon of three-dimensional black holes, and the resulting singularity is hidden by the horizon \cite{Zanelli, Zanelli2}. [Here the matter field is assumed to be a massless and conformally coupled scalar field]. Quantum mechanics has easily protected cosmic censorship. This finding can be interpreted as a natural consequence of the absence of  local conformal anomalies in three-dimensional spacetime \cite{birrell-davies}.  

\subsection{Four-dimensional trace anomaly and the semiclassical \\ Schwarzschild geometry}

In $1+3$ 
dimensions, we will work with the conformal anomaly  given by 
\be \label{4danomaly} \langle  T^{\mu}_\mu  \rangle = \hbar( c\ C^2 -a \ E) \ , \ee
where $C^2= C_{\mu\nu\rho\sigma}C^{\mu\nu\rho\sigma}$ is the square of the Weyl curvature 
and \be E=R^{\mu\nu\rho\sigma} R_{\mu\nu\rho\sigma}-4 R^{\mu\nu} R_{\mu\nu}+R^2 \  \ee
 is the Euler density. 
We have ignored contributions of the form $\Box R$ as they are intrinsically ambiguous and can be shifted by local counterterms. 
 The numerical coefficients $c$ and $a$ depend on the spin of the field. For convenience we consider here a conventional scalar field. The coefficients $a$ and $c$ are  given by \cite{birrell-davies}
\be a= \frac{1}{360(4\pi)^2} \ \ \ \ \ \ 
  c= \frac{1}{120(4\pi)^2} \ . \ee

Although the  conformal anomaly is not sufficient to determine the full expression for the renormalized stress-energy tensor $\langle T_{\mu\nu}\rangle $, with 
the reasonable approximation of a perfect fluid form for  the vacuum stress-energy tensor, 
one can adequately  estimate the backreacted static and spherically symmetric metric \cite{Pau}. 
This simplifying assumption is inspired by  the result (\ref{Tcandelas})  in the fixed Schwarzschild background when $r\to 2M$. The main features of the backreacted Schwarzschild geometry can be expected to be captured by this near-horizon approximation.

We assume now a spherically symmetric and static metric 
\be ds^2 = -e^{-2\phi(r)} dt^2 + \frac{dr^2}{1-\frac{2m(r)}{r}} +r^2 d\Omega^2 \, . \label{schwansatz}\ee
With this ansatz, the semiclassical Einstein equations are mathematically equivalent to the set of equations 
\bea 
\frac{d m(r)}{dr}&=& 4\pi r^2\langle \rho (r)\rangle \, , \label{eqnmass}\\
\frac{d \phi(r)}{dr}&=&-\frac{m(r) +4\pi r^3 \langle p_r (r)\rangle }{r^2(1- \frac{2m(r)}{r})} \label{eqnphi} \, ,\\ 
\frac{d \langle p_r (r) \rangle }{dr}&=&-\frac{m(r)+4\pi r^3\langle p_r(r)\rangle }{r^2(1- \frac{2m(r)}{r})}(\langle \rho(r) \rangle  +\langle p_r(r) \rangle )-\frac{2}{r}(\langle p_r(r) \rangle  -\langle p_t (r) \rangle) \ , \ \ \ \ \ \ \ \  \label{eqpressure}
\eea
together with an effective equation of state for $\langle T_{\mu\nu} \rangle$ given by the trace anomaly
\be -\langle \rho \rangle +\langle p_r \rangle +2\langle p_t \rangle= \langle T^\mu_\mu \rangle\, .  \label{stateeq}\ee
$\langle p_r\rangle$ and $\langle p_t\rangle$ are the mean values of the radial and tangential pressures of the vacuum stress-energy tensor $\langle T_{\mu\nu}\rangle$, respectively. We now introduce a simplifying assumption, namely $\langle p_r\rangle = \langle p_t\rangle$. It is inspired by the near-horizon behavior of $\langle T_{\mu\nu}\rangle $ obtained for the classical Schwarzschild geometry. With this input, the above semiclassical equations  have the form of the  Tolman-Oppenheimer-Volkoff (TOV) equations of classical general relativity. Taking into account the specific form of the trace anomaly (\ref{4danomaly}) one can rewrite (\ref{stateeq}) as 
\be
-\langle\rho\rangle+3\langle p_r\rangle=\frac{\hbar}{270}\left[\left(\frac{3 m}{ 4\pi  r^3}-\langle\rho\rangle \right)^2+6 \left(\langle\rho\rangle^2+3\langle p_r\rangle^2\right)-2 (3\langle p_r\rangle -\langle\rho\rangle)^2\right]\, .
\ee

 It is interesting to see the approximated form of the metric around the classical horizon $r=2M$ . The perturbative solution in powers of $\hbar$ gives \cite{Pau}: 
\bea
ds^2&=&-[1-\frac{2M}{r}-\frac{\hbar r}{13440\pi M^2 (r-2M)}+\cdots] dt^2 \nonumber \\ 
&+& [1-\frac{2M}{r}- \frac{\hbar r}{4480\pi M^2 (r-2M)}+\cdots ]^{-1} dr^2 + r^2 d\Omega^2 \, . \label{analyticalresult}
\eea
The classical Schwarzschild coordinate singularity at $r= 2M$ get shifted to the value $r_0=2M + \mathcal{O}(l_P)$, defined by $g_{rr}^{-1}(r_0)=0$. $l_P\sim \hbar^{1/2}$ is the Planck length.  However, unlike for the classical Schwarzschild geometry,  the component $g_{tt}$ of the metric does not vanish at $r=r_0$. This implies that the static spacetime we have obtained is horizonless and does not define a black hole geometry. We can also solve the exact semiclassical Einstein equations numerically, without introducing the ansatz of a perturbative power series for the metric.  
The numerical results confirm \cite{Pau}, in the region $r \approx 2M$, the form of the metric given by (\ref{analyticalresult}), up to the  numerical coefficients multiplying the factor $M^2(r-2M)$.

\begin{figure}  \begin{center}\includegraphics[angle=0, width=60mm]{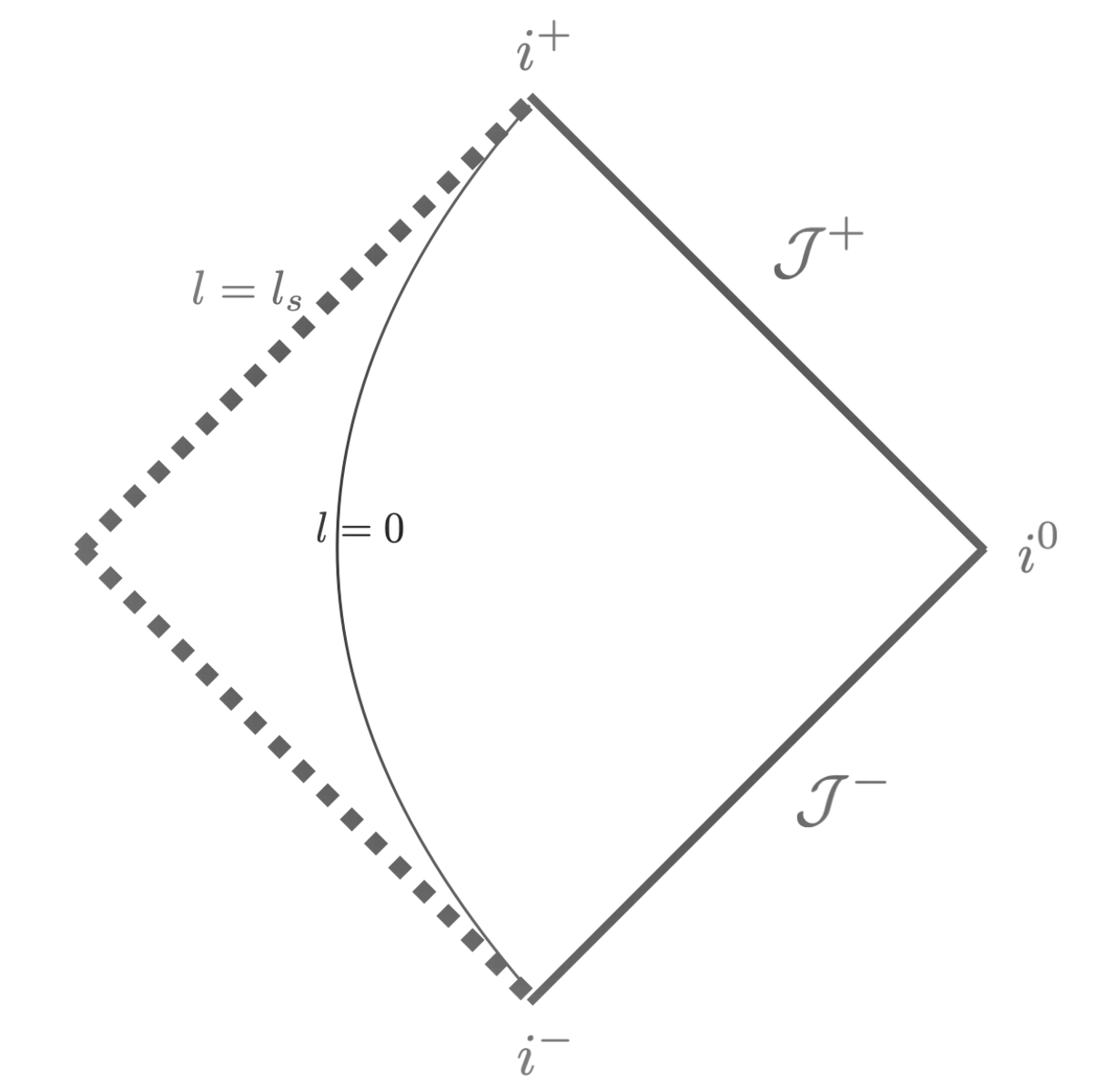}\end{center}
\caption{ Penrose diagram showing the conformal structure of the semiclassical Schwarzschild geometry obtained from the four-dimensional trace anomaly. The bouncing surface for the radial function $r(l)$ is located at $l=0$ and represents the throat of an asymmetric  wormhole.  A null curvature singularity is  located at $l=l_s<0$.
}\label{censorship2}\end{figure}

Furthermore,  the metric can be analytically extended beyond the bouncing point for the radial function at $r=r_0$. One finds a null curvature singularity immediately after crossing the wormhole throat. Introducing the  proper-length coordinate $l(r)$  the metric can be rewritten to fit the Morris-Thorne ansatz
\be 
d s^2=-e^{-2 \phi(l)} d t^2+d l^2+r^2(l) d \Omega^2 \ . \ee
The $l$ coordinate can be adjusted to locate the  throat at $l(r_0)=0$.  The curvature singularity is in the inner region, $l_s <0$, at the vanishing of the redshift function $g_{t t}\left(l_s\right)=0$, and very close to the throat  $l_s \sim \mathcal{O}(\sqrt{l_P M})$ \footnote{The location of the curvature singularity can be seen  as a side effect of the assumption of a pure vacuum solution. 
The presence of matter   could relocate  the singularity beyond the throat, thus allowing the formation of exotic compact objects (ECOs).}. 
Other  computational techniques \cite{Julio2} agree  with this description, which then appears as a very robust prediction. The near-horizon geometry is also consistent with  results obtained via the effective action \cite{BSS}. Figure \ref{censorship2} provides the corresponding Penrose diagram. 
We find the same
 overall picture  obtained by the  effective two-dimensional anomaly (\ref{2danomaly}), as depicted in Figure \ref{censorship}.

 In summary, thanks to the non-vanishing trace anomaly of quantized fields, 
 the quantum backreacted Schwarzschild  geometry is horizonless with a curvature singularity.  Cosmic censorship does not seem to be  respected by quantum mechanics. The main novelty of this work is to provide  a resolution of this tension.  This is the focus of the next section.

\section{Cosmic censorship, conformal symmetry, and the Standard Model}\label{Tcancellation}

If cosmic censorship is a deep property of nature, and not just a property of generic gravitational collapse in classical general relativity, we can go further and explore its consequences at the quantum level. 
According to the previous discussion, the conformal anomaly seems to drive the breaking of cosmic censorship. Therefore, maintaining cosmic censorship requires that the total conformal anomaly of all  fundamental fields be exactly canceled out 

\be  \label{ff0}\sum_{\rm{fundamental \ fields}} \langle T^\mu_\mu \rangle =0 \ . \ee 

 The covariant conservation and tracelessness of the total stress-energy tensor implies the following near-horizon behavior for a fixed Schwarzschild background
\be  \label{ff00}\sum_{\rm{fundamental \ fields}} \langle T^\mu_\nu \rangle \sim_{r \to 2M} \left(\begin{array}{cccc}
-\langle \rho \rangle 
 & 0 & 0 & 0 \\
0 &\langle p_r\rangle = -\langle \rho \rangle   & 0 & 0 \\
0 & 0 & \langle p_t\rangle= \langle \rho \rangle & 0 \\
0 & 0 & 0 & \langle p_t\rangle= \langle \rho \rangle \end{array}\right)
 \ . \ee 
This result can be derived with the help of  the general arguments given in \cite{CF}. 

We  can now solve the semiclassical equations expressed in the form of TOV-type equations with the equation of state 
\be -\langle \rho \rangle +\langle p_r \rangle +2\langle p_t \rangle= 0 \, ,  \label{stateeq2}\ee
and the simplifying assumption $\langle p_t\rangle = - \langle p_r\rangle$, as suggested by the near-horizon behavior (\ref{ff00}). In this situation, the semiclassical equations can be solved analytically. The result is the rescue of the event horizon. The solution takes the form

\bea
ds^2&=&-(1-\frac{2M}{r}+ \frac{Q^2}{r^2}) dt^2 
+ \frac{dr^2}{(1-\frac{2M}{r}+ \frac{Q^2}{r^2})} + r^2 d\Omega^2 \, . \label{analyticalresult2}
\eea
where the parameter $Q^2$ is proportional to the square of the Planck length, i.e.,  $Q^2 \propto l_P^2 \ll M^2$. It is mathematically similar to the classical Reissner-Norstr\"om solution for spherically charged black holes. The event horizon is  at $r_h= M + \sqrt{M^2 -Q^2}$ and the singularity is hidden to a distant observer.

Therefore, with (\ref{ff0}) one can recover
 the event horizon  in a manner consistent with backreaction and stationarity. 
 The price (or the reward) is a strong constraint on the set of fundamental fields. Let us analyze with more detail the condition (\ref{ff0}). 
The contribution of the known fields of the Standard Model to $\langle T^\mu_\mu \rangle$ (ignoring masses and interactions) is given by (\ref{4danomaly}) with the following numerical coefficients \cite{birrell-davies}
\be a= \frac{1}{360(4\pi)^2} [ N_0+ \frac{11}{2} N_{1/2} + 62 N_1 ] \ >0 \ , \ \ \ \  
  c= \frac{1}{120(4\pi)^2} [ N_0+ 3N_{1/2} + 12 N_1] \ > 0 \ , \ee
with $N_1= 12$ (electroweak bosons and gluons), $N_{1/2}= 3 \times 15$ (three generations of left-handed and right-handed leptons and quarks), $N_0= 4$ (real components of the Higgs doubled). As free fields in curved spacetime, their contribution to the conformal anomaly is always  additive and cannot be forced to cancel.


However, in sharp contrast to two-dimensional conformal invariance, it is now possible to introduce a new field with negative contribution to $c$ and $a$ while preserving unitarity. The simplest way is provided by the so-called ``dimensionless scalar field'' \cite{Bogolubov}. It is a real scalar field $\xi$ which, in flat spacetime, obeys the 4th order equation $\Box^2 \xi =0$. The real part of the two-point function is given by 
\be \langle \xi(x)\xi(y) \rangle = -(4\pi)^{-1} \log |\kappa^2 (x-y)^2|\ ,  \ee where $\kappa $ is an infrared cutoff. An important property of this  theory is that its physical content consists of a single quantum state: the vacuum \cite{Bogolubov}. The reason for this is the underlying gauge symmetry of the theory under the local transformations
$\xi(x) \to \xi(x) + \alpha(x)$, with $\Box \alpha =0$.
In curved spacetime, it can be uniquely extended to a classically invariant theory under  (\ref{1})
\be  S= -\frac{1}{2} \int d^4x \sqrt{-g} \ \xi \triangle_4 \ \xi \ ,  \ee 
 where  $\triangle_4$ is the unique conformally-invariant fourth order operator
 \be  \triangle_4 = \Box^2 + 2 R^{\mu\nu}\nabla_\mu \nabla_\nu - \frac{2}{3}R \Box + \frac{1}{3}(\nabla^\mu R) \nabla_\mu \ . \ee
A nice feature of this field is that it contributes negatively 
to the conformal anomaly
\be a= -\frac{28}{360(4\pi)^2}, \  \ \ \ \ c= -\frac{8}{120(4\pi)^2} \ . \ee
This  result can be obtained from the DeWitt-Seeley-Gilkey coefficients of the corresponding conformally invariant fourth order operator  \cite{Gusynin89}.  It opens the door for canceling the entire conformal anomaly.

Finding a solution for (\ref{ff0}) is not so easy. The problem is similar to the four non-trivial constraints for the cancellation of the gauge and gravitational anomalies in the Standard Model \cite{Schwartz} for each generation. Here, however, the problem concerns  the three generations of leptons and quarks simultaneously. Fortunately, the cancellation of the conformal anomaly 
\bea \label{ac0}a&=& \frac{1}{360(4\pi)^2} [ N_0+ \frac{11}{2} N_{1/2} + 62 N_1 -28 N^{\xi}] \ =0 \  \nonumber \\ 
  c&=& \frac{1}{120(4\pi)^2} [ N_0+ 3N_{1/2} + 12 N_1 -8 N^{\xi}] \ =0 \ , \eea
has a solution \cite{BT21, MVZ}. It requires the addition of  a 
number $N^{\xi}=36$ of ``dimensionless scalar fields''\footnote{The number itself is not very important because these fields  only contribute to the vacuum. There are no particle excitations associated with the $\xi$ fields.}, and  also three right-handed neutrinos to get $N_{1/2}=48$ ($16$ Weyl fields  for each generation) with $N_1=12$ (for the $SU(3)\times SU(2)\times U(1)$ gauge group) . 
$N_0=0$, so the conventional scalars (such as the Higgs doublet) are not allowed by the constraints (\ref{ac0}), and therefore cannot be considered as fundamental entities. 

The necessary inclusion of three flavours of sterile right-handed neutrinos reinforces the role of the heavier one  as a dark matter candidate \cite{Drewes13, Boyle}. 
 We will return to this point later.

 \section{Conformal symmetry and the Big Bang singularity}
As noted at the  beginning,  the smoothness of the initial  singularity can be protected by imposing  the WCH, in the much more precise formulation  given in \cite{Tod03, Tod09}. The Big Bang singularity is assumed to be purely conformal, in the sense that it can be reabsorbed by a conformal transformation (\ref{1}) of the metric tensor. The singularity 
is then due to the choice of metric in the conformal class. 

At the quantum level the conformal  regularity  in the stress-energy tensor of matter fields is removed by the trace anomaly. The initial singularity cannot be  fully absorbed in $\langle T_{\mu\nu} \rangle$ by a conformal transformation (see, for instance, \cite{DNT}). While the metric is conformally regular, the rescaled quantum stress-energy tensor 
\ $\Omega^{-2} \langle T_{\mu \nu} \rangle $ 
is not longer regular at the Big Bang  due to the conformal anomaly.
 This is somewhat in tension with the  WCH, the resolution of which requires again exact conformal symmetry (\ref{ff0}). We then reach  the same conclusion as in the case of black holes and cosmic censorship.

The simplest way to enforce the WCH in semiclassical gravity is by assuming a total vacuum stress-energy tensor  $\langle T_{\mu\nu}\rangle$ of a perfect fluid form  with  ``equation of state'' corresponding to pure radiation $\langle \rho\rangle= \frac{1}{3} \langle p \rangle$, i.e.,  our fundamental constraint (\ref{ff0}). 
As proven in \cite{Newman93} in a purely classical context, the above conditions for a stress-energy tensor which evolves from a spacelike conformal singularity, having a vanishing Weyl tensor 
\be \label{Weyl} C_{\mu\nu\rho\sigma} = 0 \ , \ee 
is necessarily a Friedmann-Lema\^{i}tre-Robertson-Walker (FLRW) spacetime near the singularity. This means that, near the conformal singularity
\be \label{FLRWrd0}ds^2 \sim  a^2(\eta)(-d\eta^2 + h_{ij}dx^idx^j) \ , \ee 
with $\eta$ the conformal time and \be \label{FLRWrd1}a(\eta) \propto \eta \ . \ee
This is fully consistent with observations \cite{Planck1, Planck2} if one additionally assumes that $h_{ij}$ is a spatially flat metric.

 \subsection{Gravitational particle production}
 
 Let us go back to the question of the sterile right-handed neutrinos. So far, we have been neglecting the masses and the interactions of the basic constituents. 
 The first important correction is the consideration of the mass of the heavier field, which should be one of the right-handed neutrinos. 
 According to the seesaw mechanism, the  mass matrix of the neutrinos, after symmetry breaking, is given by \cite{Schwartz}
  \begin{equation}
\mathcal{M}=\left(\begin{array}{cc}
0 & m_D \\
m_D & M_R
\end{array}\right)
\end{equation} 
 where $m_D$ is the Dirac mass 
 and $M_R$ is the intrinsic Majorana mass of the right-handed neutrino $\nu_R$. It is assumed that $m_D \ll M_R$ and thus the mass of the left-handed neutrino $\sim m_D^2/M_R$ becomes very small. Therefore, the dominant contribution is the Majorana mass $M_R$, which is assumed to be several orders of magnitude beyond the electroweak scale.  
 If, as expected, the heavier neutrino  interacts almost exclusively with gravity, the natural production mechanism is  gravitational particle creation by the early expansion of the universe 
\cite{parkerthesis, Parker68, Parker69, Parker71}. 

The mechanism involves  a well-defined quantum vacuum (i.e., a so-called Hadamard state \cite{Wald}) around the  Big Bang event.  
 Remarkably, the FLRW geometry (\ref{FLRWrd0}) provides a natural arena for the construction of  Hadamard states  around the Big Bang.  
 The so-called ``smeared'' states of low energy  \cite{Olbermann, NNP23} solve the problem. They minimize the averaged energy density $\mathcal{E}_k[f]$ over a temporal window function $f^2(\eta)$ 
\be
\label{eq:smearedED2}
\mathcal{E}_k[f] = \int \text{d} \eta\, \sqrt{|g|}\, f^2(\eta) \, \rho_k(\eta) \, ,
\ee
where  $\rho_k$ corresponds to the  energy density in momentum space 
 \be \langle\rho\rangle= \frac{1}{2 \pi^2 a^3} \int_0^{\infty} d k k^2 \rho_k(\eta) \ . \ee 
 The smearing can be done with a Gaussian function \cite{BNNP}
  \be
 f^2=\frac{1}{ \sqrt{\pi}\epsilon}e^{-\frac{\eta^2}{\epsilon^2}}\, .
 \ee
 This extra dependence of the vacuum state can be naturally interpreted as a phenomenological side effect of quantum gravity, since we are using quantum field theory and semiclassical gravity in the vicinity of the Big Bang.  The particle production spectrum at late times can then be accurately calculated. The number density of created particles for an expansion factor of the form (\ref{FLRWrd1}), and a spatially-flat metric $ds^2 \sim \eta^2(-d\eta^2 + d\vec x^2)$, is given by \cite{BNNP}
 \be \label{ntotal2}\langle n(t) \rangle= \alpha(\epsilon) \left(\frac{M_R}{t} \right)^{3/2} \ , \ee
 where $t$ is proper time and the numerical coefficient $\alpha(\epsilon)$ keeps the dependence on the window function. One can estimate the mass $M_R$  to fit the dark matter abundance at the end of the radiation-dominated epoch. In the limit $\epsilon \to 0$ one finds  $M _R(\epsilon=0)\approx 3 \times 10^{8}$ GeV, while in the opposite limit one gets  $M_R(\epsilon=\infty)\approx 5 \times 10^{8}$ GeV \footnote{Numerical estimations for the mass of a very weakly interacting superheavy boson give $m \sim  10^{13}$ GeV \cite{chung-kolb-riotto}.}.   The second limit corresponds to a very wide window function and turns out to be equivalent to the proposal given in \cite{Boyle}.
   When the window function is essentially supported at the Big Bang (i.e., $\epsilon \to 0$), the physical momenta $\vec{k}/a(\eta)$ are extremely blue-shifted and the field has  a vanishing effective mass $M_R^2 \ll \vec{k}^2/a^2(\eta)$. A massless spin-$1/2$ field is conformally invariant and it has a natural vacuum state, called the conformal vacuum.  
 Therefore, the limit $\epsilon \to 0$ mimics the conformal vacuum.

\section{Conclusions and final comments}
 
 Quantum field theory, via gravitational trace anomalies, creates tension with Penrose's cosmic censorship and  the WCH. The requirement of exact conformal symmetry (anomaly cancellation) removes both tensions and has important implications for constraining the set of fundamental fields.
There are fundamental observations 
that indirectly support this  view.
 One is the recent LIGO-Virgo detections \cite{LIGO-Virgo} of binary black hole mergers, which are consistent with the non-trivial fact that the resulting objects behave as larger black holes rather than as naked singularities (also consistent with the predictions of numerical relativity \cite{Pretorius}). 
  Furthermore, observations also indicate that soon after the Big Bang, the Universe is well described, up to tiny perturbations, by an  extremely homogeneous and isotropic spacetime dominated by purely classical radiation 
\cite{Planck1, Planck2},  
which is the easiest  way to have a vanishing conformal anomaly taming the initial singularity. 

 The  discussion of  the calculation of the predicted mass of the heaviest right-handed neutrino   illustrates the need to go beyond pure quantum field theory to resolve the intrinsic ambiguities of field theory.
 However, the overall picture underlying this work strongly suggsts that field theory and semiclassical gravity, when guided by symmetry principles,  can provide deep insights. 
 

Finally, we should also remark that in the above discussion we have neglected the interactions between the fundamental fields, which should also contribute to the conformal anomaly through running coupling constants and beta functions. A natural extension of the above arguments is to constrain the fundamental interactions to ensure exact conformal symmetry. A  similar proposal in this direction has been made in \cite{Hooft}. In any case, it is very suggestive that such a local symmetry protects both cosmic censorship and the WCH 
by involving  the field content of the Standard Model. \\

{\bf Acknowledgements.} 
The content of this work
 has been presented 
 at the ``Fourth European Physical Society Conference on Gravitation: Black Holes'' (Valencia, Spain; 13-15 November 2023). The author is grateful to  the audience  for  positive discussions and  feedback. 
This work is  supported by the Spanish Grant PID2020-116567GB- C21 funded by MCIN/AEI/10.13039/501100011033 and the project PROMETEO/2020/079 (Generalitat Valenciana).



\begin{thebibliography}{99}

\bibitem{Penrose65} Penrose R 1965 {\it Phys. Rev. Lett.} {\bf 14} 57 

\bibitem{HP70}  Hawking S W and Penrose R 1970 {\it Proc. Roy. Soc. Lond. A} {\bf 314}, 529 


\bibitem{Penrose69}  Penrose R 1969 {\it Riv. Nuevo Cim.} {\bf 1}, 257;  {\it Gen.Rel.Grav.} {\bf  34} (2002) 1141-1165. 

\bibitem{Penrose73}  Penrose R 1973  {\it Ann. N.Y. Acad. Sci.} {\bf 224}, 125 

\bibitem{Penrose79} Penrose R 1979, ``Singularities and time-asymmetry''. In {\it General Relativity: an Einstein Centenary}, ed. Hawking S W and  Israel W, Cambridge University Press, Cambridge 

\bibitem{Tod03} Tod K P 2003 {\it Class. Quant. Grav. } {\bf 20}, 521

\bibitem{Schwartz}  Schwartz M D 2014 {\it Quantum Field Theory and the Standard Model}, Cambridge University Press, Cambridge 

\bibitem{CD}    Capper D M and Duff M J, 1974
        {\it Nuovo Cim. A} {\bf 23} 173



\bibitem{DDI} Deser S Duff M and Isham C J  1976 {\it Nucl. Phys .B} {\bf  111} 45 


\bibitem{birrell-davies} Birrell N D  and Davies P C W 1982 {\it Quantum Fields in Curved Space}, Cambridge University Press, Cambridge, England 


\bibitem{parker-toms} Parker L and Toms D J 2009 {\it Quantum Field Theory in Curved Spacetime: Quantized Fields
and Gravity}, Cambridge University Press, Cambridge, England

\bibitem{hu-verdaguer} Hu B-L B and Verdaguer E 2020 {\it Semiclassical and Stochastic Gravity: Quantum Field Effects on Curved Spacetime},  Cambridge  University Press, Cambridge, England (2020).

\bibitem{Duff} Duff M J 1994 {\it Class. Quant. Grav.} {\bf 11} 1387 


\bibitem{Cardoso-Pani} Cardoso V and Pani 2019, {\it  Living Rev. Rel. } {\bf 22}, 1


\bibitem{BT21}  Boyle L and Turok N 2021 ``Cancelling the vacuum energy and Weyl anomaly in the standard model with dimension-zero scalar fields'', arXiv:2110.06258. 

\bibitem{MVZ} Miller J  Volovik G E and  Zubkov  M A 2022 {\it Phys. Rev. D} {\bf 106} 015021 



\bibitem{Minkowski} Minkowski P 1977 {\it Phys. Lett. B} {\bf 67} 421

\bibitem{SM} Burguess C and Moore G 2006 {\it The Standard Model: a Primer},  Cambridge University Press, Cambridge, England

\bibitem{parkerthesis}  Parker L 1966 {\it The creation of particles in an expanding universe}, Ph.D. thesis, Harvard University 

\bibitem{Parker68} 
   Parker L 1968
  {\it Phys. Rev. Lett.}  {\bf 21} 562
  
\bibitem{Parker69} 
  Parker L 1969
  {\it Phys. Rev.}  {\bf 183} 1057
  
\bibitem{Parker71} 
  Parker L 1971
  {\it Phys. Rev. D} {\bf 3} 346


\bibitem{Olbermann}   Olbermann H 2007 {\it Class. Quant. Grav.} {\bf 24} 5011 

\bibitem{BN} Banerjee R and Niedermaier M 2020 {\it J. Math. Phys.} {\bf 61}, 103511 

\bibitem{NNP23} Nadal-Gisbert S  Navarro-Salas J and  Pla S 2023 {\it Phys. Rev. D} {\bf 107} 085018 

\bibitem{Hooft}  't Hooft G 2015 {\it Int. J. Mod. Phys. D} {\bf 24}, 1543001 

\bibitem{Candelas80} Candelas P 1980 {\it Phys. Rev. D} {\bf 21} 2185

\bibitem{Boulware} Boulware D G 1975 {\it  Phys. Rev. D} {\bf 11} 1404 

\bibitem{Howard-Candelas} K. W. Howard K W and Candelas P 1984 {\it  Phys. Rev. Lett.} {\bf 53} 403

\bibitem{Anderson-Hiscock-Samuel} Anderson P R Hiscock W A  and Samuel D A 1995 {\it Phys.
Rev. D} {\bf 51} 4337 

\bibitem{Levi-Ori} Levi A and Ori A 2015 {\it Phys. Rev. D} {\bf 91}  104028 

\bibitem{TBO} Taylor P Breen C and Ottewill A 2022 {\it  Phys. Rev. D} {\bf 106}
065023 

\bibitem{Harvey}  Harvey J A and Strominger A 1992, ``Quantum Aspects of Black Holes'', hep-th/9209055. 

\bibitem{Strominger} Strominger A 1995, ``Les Houches Lectures on Black Holes'', hep-th/9501071.

\bibitem{Mukhanov94}  Mukhanov V  Wipf A and Zelnikov A 1994 {\it Phys. Lett. }{\bf 332} 283 

\bibitem{FN05} Fabbri A and Navarro-Salas J 2005 {\it Modeling black hole evaporation}, ICP-World Scientific, London 

\bibitem{CFT} Philippe F Mathieu P Sénéchal D 1996 {\it Conformal Field Theory}, Springer-Verlag, New-York


\bibitem{FN06}  Fabbri A   Farese S  Navarro-Salas J  Olmo G and  Sanchis-Alepuz H 2006 {\it Phys. Rev. D} {\bf 73} 104023 


\bibitem{Julio20}   Arrechea J  Barcel\'o C  Carballo-Rubio R and Garay L J 2020 {\it Phys. Rev. D} {\bf 101} 064059 

\bibitem{Ho}   Ho P M and Matsuo Y 2018  {\it Class. Quantum Grav.} {\bf35} 065012 

\bibitem{FQS84}  Friedan D Qiu Z and  Shenker S 1984 {\it Phys. Rev. Lett.} {\bf 52} 1575 

\bibitem{P} Polyakov A M 1981    {\it Phys. Lett. B} {\bf 103} 207








\bibitem{Zanelli}  Casals M A. Fabbri A  Mat\'inez C and  Zanelli J 2017 {\it Phys. Rev. Lett.} {\bf 118} 131102 

\bibitem{Zanelli2}  Casals M A. Fabbri A  Mat\'inez C and  Zanelli J 2019 {\it Phys. Rev. D} {\bf 99} 104023 




\bibitem{Pau}  Beltr\'an-Palau P   del R\'io A and  Navarro-Salas J 2023    {\it  Phys. Rev. D} {\bf 107} 085023 

\bibitem{Julio2}   Arrechea J  Barcel\'o C  Carballo-Rubio R and Garay L J 2023 {\it  Phys. Rev. D} {\bf 107} 085023 

\bibitem{BSS} Berthiere C Sarkar D and Solodukhin S N 2018 {\it Phys. Lett. B} {\bf 786}  21

\bibitem{CF} Christensen S M and Fulling S A 1977 {\it Phys. Rev. D} {\bf 15} 2088

  


\bibitem{Bogolubov}  Bogolubov N N Logunov A A  Oksak A I and  Todorov I T 1990, {\it General Principles of Quantum Field Theory}, Kluwer Academic Publishers, Dordrecht 


\bibitem{Gusynin89}  Gusynin V P 1989 {\it Phys. Lett. B} {\bf 225} 233 

 \bibitem{Drewes13} Drewes M  2013   {\it Int. J. Mod. Phys. E} {\bf 22}, 1330019
\bibitem{Boyle} Boyle L  Finn K and Turok N 2018     {\it Phys. Rev. Lett.} {\bf 121} 251301 

\bibitem{Tod09} Tod P 2010 {\it J. Phys.: Conf. Ser.} {\bf 229} 012013

\bibitem{DNT} del R\'io A Ferreiro A Navarro-Salas J and Torrent\'i F 2017 {\it Phys. Rev. D} {\bf 95} 105003
    
\bibitem{Newman93} Newman R P A C 1993 {\it Proc. Roy. Soc. Lond. A}
{\bf 443} 473; {\it Proc. Roy. Soc. Lond. A} {\bf 443} 493 

\bibitem{Planck1}  Ade P A R {\it et al.} 2016 [Planck Collaboration], 
{\it Astron. Astrophysics.} {\bf 594}  A13 

\bibitem{Planck2} Aghanim N et al. 2020 [Planck collaboration] {\it 
Astron. Astrophys.} {\bf  641}, A1 



\bibitem{Wald}  Wald R M 1995 {\it Quantum Field Theory in Curved Space-Time and Black Hole Thermodynamics}, University of Chicago Press, Chicago 
 
\bibitem{BNNP}  Beltr\'an-Palau P Nadal-Gisbert S Navarro-Salas J Pla S   2023 {\it J. Phys. Conf. Ser.} {\bf 2531}  012009





\bibitem{chung-kolb-riotto} Chung D J H Kolb E W and Riotto A 1998 {\it Phys. Rev. D} {\bf 59}, 023501 


\bibitem{LIGO-Virgo} Abbott B P {\it et al.} 2016 (LIGO Scientific Collaboration and Virgo Collaboration), {\it Phys. Rev. Lett.} {\bf 116} 061102; {\it Phys. Rev. Lett.} {\bf 116} 241103 

\bibitem{Pretorius} Pretorius F 2005 {\it Phys. Rev. Lett.} {\bf 95} 121101





   
\end{thebibliography}
\end{document}